\documentstyle[preprint,prc,aps]{revtex} 
\begin{document}
\draft
\title{Reply to Comment on ``Determination of pion-baryon coupling\\
constants from QCD sum rules"}
\author{Boris Krippa$^a$, Michael C. Birse$^b$}
\address{$^{a}$Institute for Nuclear Research of the
Russian Academy of Sciences, Moscow 117312, Russia\\
$^{b}$Theoretical Physics Group, Department of Physics and Astronomy 
University of Manchester, Manchester, M13 9PL, UK\\}
\maketitle
\vskip1cm

In his Comment\cite{Kim}, Kim criticises the sum rules obtained in 
Ref.\cite{BK} for pion-baryon coupling constants. In particular he suggests
that the treatment of the continuum in our work is inconsistent, and he 
presents a different perturbative model for the continuum that leads to rather 
different results.

However Kim's arguments rely heavily on the use of single dispersion relations 
to take into account the continuum contributions to the correlation function. 
These are assumed to take the form
\begin{eqnarray}
\Pi^{\rm cont}(p^2) = \int^\infty_S ds {\rho_{\rm OPE}(s)\over s-p^2}+\cdots,
\end{eqnarray}
up to subtraction terms which are needed to cancel divergences of the integral
over $s$ and which form a polynomial in $p^2$. Here is $S$ is some threshold
above which the perturbative continuum from the operator-product expansion
(OPE) is assumed to provide a good approximation to the true spectral density
$\rho(s)$. This is the form familiar in the analysis of sum rules for hadron
masses, which are obtained from vacuum-to-vacuum two-point correlators. In
contrast, meson-to-vacuum correlators or correlators in the presence of an
external field should be represented by a double dispersion relation. This is
because the external meson or field can cause transitions between different
states that are created or annihilated by the chosen interpolating fields. A
separate dispersion relation is thus needed for each of the hadron
``propagators" corresponding to the initial and final hadrons. (For more
details, see Ref.\cite{Iof95}). In the general case where the external meson
or field carries nonzero momentum $q$, the continuum contribution can be
written as
\begin{eqnarray}\label{2dr}
\Pi^{\rm cont}(p^2_1,p^2_2,q^2)= \int^\infty_{S_1} ds_1 
\int^\infty_{S_2} ds_2 
{\rho_{\rm OPE}(s_1,s_2,q^2) \over (s_1 - p_1^2)(s_2 - p_2^2)}+\cdots,
\end{eqnarray}
where $S_1$ and $S_2$ are (possibly different) thresholds. 

To obtain the sum rules of Ref.\cite{BK} we expanded around the chiral limit
and so we worked in the limit $q^2\rightarrow 0$. In this case $p_1^2=p_2^2
\equiv p^2$ and so a single momentum flows through the correlator. At first 
sight one might think that the use of the single dispersion relation in $p^2$ 
is legitimate, since one could split up the denominator in (\ref{2dr}) and 
perform the integration over either $s_1$ or $s_2$ first. However, this
impression is misleading: the integrals over $s_1$ and $s_2$ are ultraviolet
divergent and, moreover, the subtraction terms that are needed for the
integral over $s_1$ ($s_2$) appear multiplying an unknown function of $p_2^2$
($p_1^2$). Hence, even in the limit of $q^2\rightarrow 0$, one cannot cancel
the divergences of the integral in Eq.~(\ref{2dr}) by a simple polynomial in
$p^2$. The above representation of $\Pi^{\rm cont}(p^2,p^2,0)$ is thus not
equivalent to a single dispersion relation. As stressed by Ioffe\cite{Iof95},
important features of QCD sum rules for coupling constants, such as the double
pole at $p^2=M_N^2$ and the single pole due to nucleon-to-continuum
transitions and the associated subtraction terms, rely on the use of a double
dispersion relation. 

As an example of our treatment of the continuum, consider the dimension-5 term 
in our sum rule\cite{BK}, which arises from a term of the form $C\ln(-p^2)$ in 
the OPE. The corresponding spectral density that reproduces this logarithm 
(up to subtraction constants) has the form
\begin{equation}
\rho_{\rm OPE}(s_1,s_2,0)=-Cs_1\delta(s_1 - s_2).
\end{equation}
If we use this perturbative density in Eq.~(\ref{2dr}), starting at some 
threshold $S$, as our model for the continuum on the phenomenological side of 
the sum rule, we obtain
\begin{equation}\label{pcont5}
\Pi^{\rm cont}(p^2,p^2,0)=-C\int_S^\infty ds {s\over (s-p^2)^2}
=C\left[\ln(S-p^2)-{S\over S-p^2}+\cdots\right],
\end{equation}
up to terms that vanish after Borel transforming.

Taking the Borel transform of Eq.~(\ref{pcont5}) with respect to $Q^2=-p^2$ 
gives a perturbative continuum contribution of the form
$$-CM^2\left(1+{S\over M^2}\right)\exp(-S/M^2).$$
When this is taken over to the OPE side of the sum rule, it leads to the
replacement of $-CM^2$ (the Borel transform of $C\ln(-p^2)$) by 
$-CM^2 E_1(S/M^2)$, as in our paper\cite{BK}. (The functions $E_n(x)$ are 
defined in the usual way: $E_n(x)=1-(1+x+\cdots+x^n/n!)e^{-x}$.) A similar 
treatment of the dimension-3 term generates a factor of $E_2(S/M^2)$.

The approach outlined here, which was used in Ref.\cite{BK}, is based on 
a simple perturbative model for the spectral density, which is assumed to start
at some threshold $S$. This is then inserted in the double dispersion relation
for the correlator where it gives rise to logarithmic discontinuities, starting
at the threshold $p^2=S$, as well as threshold singularities, which in this 
case are poles. Kim raises questions about the unphysical nature of these
poles and their consistency with ideas of duality. However duality tells us 
only that the hadronic spectral density at high energies can be well 
approximated by a spectral density of quarks and gluons. The whole threshold, 
together with any associated singularities, is an artefact of our crude
modelling of the continuum at lower energies, where hadronic resonances are
important. Moreover, any simple pole-plus-continuum ansatz for the 
phenomenological spectral density ignores many of the singularities (cuts and 
threshold singularities) that will be present in the real correlator. The whole
sum-rule approach relies on the assumption that in some averaged sense the main 
features of the real correlator are reproduced by the ansatz used on the 
phenomenological side of the sum rule. Hence any model of this type for the 
continuum should be used only in the context of some procedure for averaging 
over $p^2$, such as the Borel transform. Its detailed form as a function of 
$p^2$ should not be taken too seriously. 

We therefore believe that our treatment of the continuum is consistent with
duality and, more importantly, with the fact that the correlator in the 
presence of an external meson or field should be represented by a double 
dispersion relation.

Finally we do acknowledge one correction which does need to be made to the 
results of Ref.\cite{BK}. This concerns the contribution of the dimension-7
condensate. In Ref.\cite{BK} a contribution of this term to the continuum
was included, which led to a factor of $E_0(S/M^2)$ in that term in
the sum rule. Since, as Kim points out\cite{Kim}, the corresponding term 
in the OPE has the form $1/p^2$, with no logarithm, it does not contribute to 
the perturbative continuum. The factor $E_0$ should therefore be replaced by 1.
However this term is small; indeed it was included only in order to estimate 
the size of dimension-7 contributions to the sum rule. Hence the numerical 
results of Ref.\cite{BK} remain practically unchanged.

\end{document}